\documentclass[%
superscriptaddress,
 amsmath,amssymb,
 aps, prx,
]{revtex4-2}

\usepackage{graphicx}
\usepackage{dcolumn}
\usepackage{bm}
\usepackage{hyperref}

\begin{document}

\preprint{APS/123-QED}

\title{Beyond Monolithic Scaling: Modularity and Heterogeneity as an Architectural Imperative for Utility-Scale Quantum Computing}

\author{Bo Fan}
\email{bo.fan@pku.edu.cn}
\affiliation{Yangtze River Delta Industrial Innovation Center of Quantum Science and Technology, Suzhou, 215131, China}%
\author{Renzhou Fang}%
\affiliation{Yangtze River Delta Industrial Innovation Center of Quantum Science and Technology, Suzhou, 215131, China}%


\author{Yuntao Zhang}
\affiliation{China Academy of Electronics and Information Technology, Beijing, 100041, China}%

\author{Xiaolong Yuan}
\affiliation{Yangtze River Delta Industrial Innovation Center of Quantum Science and Technology, Suzhou, 215131, China}%

\author{Dafa Zhao}
\affiliation{Yangtze River Delta Industrial Innovation Center of Quantum Science and Technology, Suzhou, 215131, China}%


\date{\today}

\begin{abstract}
Scalable quantum computing is inherently bottlenecked not by qubit count or fabrication yield, but by a rigid temporal mismatch: macroscopic classical coordination latency ($\tau_c$) inevitably grows with system diameter, while microscopic quantum coherence ($\tau_q$) remains strictly bounded. Beyond a critical scale, this mismatch breaches the classical control light cone, triggering a superlinear geometric penalty ($\epsilon > 0$) that renders monolithic synchronization structurally unstable at scale. We formalize the resulting structural phase transition through a governing scaling law, $1+\epsilon > \gamma$, which strongly biases modular decomposition and a shift from global unitaries to Local Operations and Classical Communication (LOCC). To manage the resulting resource contention under strict coherence budgets, we introduce a layered semantic architecture and a time-aware Reserve--Commit protocol. By embedding predictive temporal pre-validation, the protocol acts as an architectural semantic classifier: it preemptively aborts transactions that exceed the causal horizon and explicitly converts scheduling-induced failures into location-known erasure metadata, directly relaxing hardware fidelity thresholds for downstream QEC decoders. Contingent on near-term transduction targets ($\eta_{\mathrm{trans}} \sim 0.1$), we project a plausible crossover scale on the order of $N_c \sim 10^5$--$10^6$ physical qubits, though this boundary remains sensitive to underlying hardware and topological parameters. This threshold marks a profound architectural convergence: the footprint required for modularity aligns precisely with early fault-tolerant utility, establishing time-aware distributed orchestration, rather than monolithic expansion or centralized classical control, as the physical imperative for utility-scale quantum computing.
\end{abstract}

\maketitle

\section{\label{sec1}Introduction}
Quantum computing has progressed from proof-of-principle demonstrations toward the early stages of system integration. Small and intermediate-scale quantum processors have been realized across a diverse range of physical platforms, including superconducting circuits, trapped ions, neutral atoms, semiconductor spins, and photonic systems \cite{arute2019quantum,bluvstein2024logical,monroe2013scaling,vandersypen2017interfacing}. These advances have shifted the central challenge of the field: from the realization of isolated qubits and elementary gates to the construction of scalable, controllable, and economically viable quantum computing infrastructures.

At the level of abstract quantum computation, this physical diversity appears largely incidental. Distinct physical platforms are computationally equivalent under standard abstractions: they support universal gate sets, admit fault-tolerant constructions, and belong to the same complexity class. From this perspective, heterogeneity is often treated as a transient feature of an immature technology landscape—an engineering fragmentation that will eventually converge toward a single dominant physical implementation as fabrication matures.

This work argues that such an expectation is fundamentally misplaced. The apparent equivalence of quantum computing platforms is an artifact of abstraction. Standard computational models deliberately suppress the physical costs associated with environmental isolation, control bandwidth, and spatial information propagation. While this suppression is indispensable for defining quantum algorithms, it obscures the rigid physical constraints that become decisive once quantum computation is embedded in large-scale dynamical systems \cite{lieb1972finite,cheneau2012light,hastings2010locality}. When these suppressed costs are reintroduced, the symmetry between platforms breaks down.

At utility scale, a quantum computer ceases to be a pure quantum mechanical abstraction; it is intrinsically a macroscopic quantum-classical hybrid thermodynamic system. Within this system, information propagation is strictly mediated by two distinct physical media: bare quantum state evolution (bounded by the Lieb–Robinson velocity, $v_{\mathrm{LR}}$) and classical measurement-feedforward signaling (constrained by electromagnetic propagation $c$ and macroscopic decoding latency). Under these dual constraints, maintaining a monolithic, homogeneous architecture incurs an unavoidable geometric and causal penalty. Specifically, as the system size scales up to accommodate millions of physical qubits, its spatial footprint inevitably breaches the control light cone—the causal horizon defined by the finite propagation speed of information and global coordination signals \cite{cheneau2012light}.

Consequently, forcing a monolithic architecture to operate beyond this causal horizon results in a superlinear explosion of space–time coordination overhead (characterized by a rigid excess geometric exponent $\epsilon > 0$). Crucially, this geometric penalty is intrinsic: quantum error correction (QEC) does not eliminate it, but merely redistributes the underlying physical noise into this geometric space–time overhead.

To bypass this geometric bottleneck, architectures must embrace modular decomposition, transforming the rigid spatial constraint into a graph-theoretic network routing problem (characterized by a scaling exponent $\gamma$ determined by routing topology). As long as network topological optimization outperforms brute-force geometric scaling, modular decomposition will eventually dominate. Architectural modularity is therefore not a transitional artifact of incomplete engineering; as we will demonstrate, it represents a scaling-induced structural tendency. We posit that, under current transduction benchmarks and surface-code overhead models, there exists a critical crossover scale ($N_c \sim 10^5-10^6$ physical qubits) beyond which the asymptotic cost of homogeneous monolithic scaling is outpaced by the efficiency of modular specialization \cite{kim2023evidence}.

The central claim of this paper is that modularity and entanglement-centric networking are structural consequences of scalable quantum computation beyond $N_c$. While modularity is the structural necessity driven by spatiotemporal light cones, architectural heterogeneity is its ultimate functional consequence---enabled by the standardized LOCC interfaces to overcome the inherent isolation--interactability--topology trilemma of any single physical substrate. Designing future quantum infrastructures requires abandoning the search for a single ``perfect'' qubit and instead establishing the semantic and architectural interfaces capable of coherently integrating diverse, distributed quantum resources.

Specifically, this perspective makes three contributions. First, we formalize a spatiotemporal scaling framework yielding the asymptotic crossover condition ($1+\epsilon > \gamma$) and a quantitative boundary ($N_c \sim 10^5$--$10^6$), establishing modularity as a causally enforced structural tendency (Section~\ref{sec2}). Second, we establish LOCC and asynchronous networking as the necessary execution models (Sections~\ref{sec3}--\ref{sec4}), from which we derive a strict causal locality bound demonstrating that classical coordination logic must be physically co-located with quantum modules (Section~\ref{sec5}). Third, we introduce a layered semantic architecture and a predictive Reserve--Commit protocol that functions as an architectural semantic classifier, converting coherence-budget violations into structured erasure metadata and thereby elevating the effective QEC threshold toward the erasure regime (Sections~\ref{sec6}--\ref{sec7}). Section~\ref{sec8} synthesizes these contributions into testable co-design imperatives for the field.

Collectively, these contributions reframe finite coherence not as a passive hardware limitation to be mitigated, but as the primary architectural driver: it dictates modular boundaries, enforces control locality, and shapes protocol semantics, transforming a fundamental physical constraint into a structured scheduling budget for fault-tolerant operation.

We emphasize that this work does not propose new physical mechanisms, but rather consolidates known spatiotemporal constraints into a testable architectural scaling model and introduces a causal reservation protocol to manage them. At intermediate scales ($N_{\mathrm{phys}} \lesssim 10^5$), homogeneous monolithic architectures may remain economically and technically superior. Furthermore, our analysis is firmly bounded by known physical mechanisms. Absent a disruptive substrate that simultaneously resolves the isolation--interactability--topology trilemma, we estimate the modular crossover to reside within $N_c \sim 10^5$--$10^6$ physical qubits. The purpose of this framework is to define the system architecture and semantic interfaces required for this scaling-induced transition.

\section{\label{sec2}Modularity as a Scaling-Induced Structural Tendency}
To substantiate the inevitability of architectural modularity, this section translates core physical constraints into formal scaling laws. We trace how the microscopic tension between quantum isolation and classical control inevitably compounds into a macroscopic space–time bottleneck under finite signal propagation. By evaluating these physical limits asymptotically, we mathematically derive the phase transition boundary where homogeneous monolithic scaling collapses, rendering modular decomposition asymptotically favored.

\subsection{\label{sec2.1}Structural Tension in Large-Scale Quantum Systems}
Large-scale fault-tolerant quantum computation requires the simultaneous optimization of three incompatible physical axes:

          \begin{itemize} 
          \item \textbf{Isolation} — suppression of environmental decoherence.
          \item \textbf{Controllability} — high-fidelity gate implementation.
          \item \textbf{Connectivity} — scalable entangling operations across distant qubits.
          \end{itemize} 
          
These axes are not independently tunable. The system Hamiltonian, carrying the dimensions of energy, may be written schematically as:
          \begin{align}
           H = H_{\mathrm{sys}} + H_{\mathrm{ctrl}} + H_{\mathrm{env}}
          \end{align}
Generically, the control and system Hamiltonians do not commute:
        
          \begin{align}
           [H_{\mathrm{sys}}, H_{\mathrm{ctrl}}] \neq 0
          \end{align}
Consequently, improved controllability necessarily introduces back-action that degrades isolation. Likewise, enhanced connectivity increases interaction pathways and control cross-talk. This incompatibility is structural rather than technological; it arises directly from non-commutativity and open-system dynamics.

To illustrate how this structural tension manifests in practice, we compare representative physical platforms. A commonly used dimensionless performance proxy is the number of operations executable within a coherence window, given by the ratio of two temporal parameters:
          \begin{align}
           N_{\mathrm{ops}} \approx \frac{\tau_q}{\tau_{\mathrm{gate}}}
          \end{align}
where $\tau_q$ stands for coherence time and $\tau_{\mathrm{gate}}$ represents gate time.

\begin{table*}
\caption{\label{tab:table1}\textbf{Representative isolation-interactability trade-offs across physical modalities.} Comparison of key performance metrics across leading quantum computing platforms. $\tau_q$ represents coherence time (isolation), $\tau_{\mathrm{gate}}$ represents gate time (controllability), and $N_{\mathrm{ops}} \approx \tau_q/\tau_{\mathrm{gate}}$ represents the number of operations executable within a coherence window. No single platform achieves Pareto optimality across all axes simultaneously, motivating architectural heterogeneity~\cite{awschalom2021development}. Data represents typical experimental values under current technological conditions.}
\begin{ruledtabular}
    \begin{tabular}{lllll}
    \textbf{Platform} & $\tau_q$(Isolation) & $\tau_{\mathrm {gate}}$(Controllability) & $N_{\text {ops }}$ & \textbf{Architectural Profile }\\
\hline Superconducting & $50-500 \mu \mathrm{~s}$ & $20-100 \mathrm{~ns}$ & $10^3-10^5$ & Fast logic, but severe crosstalk at scale \cite{krantz2019quantum}. \\
       Neutral Atom (Rydberg) & $50-200 \mu \mathrm{~s}$ & $0.2-2 \mu \mathrm{~s}$ & $10^2-10^3$ & Interaction mode; limited gate depth. \\
       Neutral Atom (Hyperfine) & $1-10 \mathrm{~s}$ & N/A (Memory) & Storage & Exceptional memory; lacks direct logic. \\
       Trapped Ion & $1-100 \mathrm{~s}$ & $1-100 \mu \mathrm{~s}$ & $10^4-10^7$ & High fidelity, but slow logical clock cycle. \\
    \end{tabular}
\end{ruledtabular}
\end{table*}

At modest scales, such trade-offs can be balanced within a single modality. The central question is whether this balance remains asymptotically stable as the dimensionless count of the physical system size $N$ increases.

\subsection{\label{sec2.2}Locality and Coordination Constraints}

Architectures based on local interactions are rigidly governed by finite-speed correlation propagation. While bare quantum correlations are bounded by the Lieb--Robinson velocity ($v_{\mathrm{LR}}$) \cite{lieb1972finite,cheneau2012light,hastings2010locality}, fault-tolerant computation is additionally constrained by a macroscopic 'classical control light cone' dictated by the end-to-end latency of the classical control-and-feedback loop.

Crucially, the physical origin of this coordination latency ($\tau_c$) is strongly substrate-dependent. We formalize this causal boundary by defining a classical control graph $G_c=(V,E)$, where vertices represent qubit registers and edges denote classical feedforward links with an effective signal velocity $v_c$. In solid-state arrays (e.g., superconducting circuits), $v_c$ is primarily bounded by electromagnetic signal flight times across massive cryostats and RF wiring. Conversely, in ultra-dense neutral-atom arrays, $v_c$ is bottlenecked by the physical limits of macroscopic control peripherals---specifically, the acoustic transit delays ($v_s$) in acousto-optic deflectors (AODs) required for dynamic reconfigurations, alongside the latency of external classical CMOS/FPGA decoding pipelines.

The classical control light cone is then rigorously defined as the maximal causal subgraph satisfying $\max_{i,j \in V} d_c(i,j)/v_c < \tau_q$, where $d_c$ is the shortest-path latency including decoding and routing overhead. Breaching this bound forces the global coordination latency $\tau_c \gtrsim \tau_q$, directly triggering the superlinear geometric penalty derived below.

Under current control architectures and standard surface-code overhead models \cite{fowler2012surface}, we define the total structural space–time cost (quantified in qubit-seconds) required to sustain a computational footprint of $N$ physical qubits. To ensure strict dimensional consistency, all structural costs are evaluated per fixed algorithmic depth (or a constant number of QEC cycles). Under this normalization, the total cost decomposes into local physical overhead and global coordination overhead:
          \begin{align}
          C_{\mathrm{hom}}(N) = C_{\mathrm{phys}}(N) + C_{\mathrm{coord}}(N)
          \end{align}
The local physical overhead $C_{\mathrm{phys}}(N)$ (encompassing physical qubits, localized wiring, and cryogenic infrastructure) represents the baseline space–time volume for local operations. Since the temporal depth is fixed by normalization, this term scales directly linearly with the physical footprint:
          \begin{align}
          C_{\mathrm{phys}}(N) = \Theta(N)
          \end{align}
          
However, fault-tolerant operation requires repeated global cycles, such as syndrome extraction and classical feedforward. For a D-dimensional embedding, the characteristic linear dimension scales as $L \sim N^{1/D}$. Because the propagation velocities of both quantum correlations (bounded by $v_{\mathrm{LR}}$) and classical control signals are strictly finite, the temporal latency of each global cycle is lower-bounded as $\Omega(N^{1/D})$ \cite{kim2023evidence,google2023suppressing}. Consequently, the total coordination volume is the product of the macroscopic spatial footprint and this temporal delay:
          \begin{align}
          C_{\mathrm{coord}}(N) \sim \text{Space} \times \text{Time} \gtrsim \Theta(N) \cdot \Omega(N^{1/D}) = \Omega(N^{1+1/D})
          \end{align}
         
To generalize this scaling beyond strict $D$-dimensional nearest-neighbor assumptions while capturing the intrinsic routing and synchronization latency, we define the dimensionless excess geometric exponent $\epsilon$. Under synchronization-constrained fault-tolerant operation, this yields an effective superlinear coordination overhead:
          \begin{align}
          C_{\mathrm{coord}}(N) = \Omega(N^{1+\epsilon}), \quad \epsilon > 0
          \end{align}
where $\epsilon > 0$ encapsulates the intrinsic physical resistance of any single substrate to infinite monolithic scaling. Building upon the substrate-specific causal bottlenecks identified above, this penalty physically manifests either as topological routing congestion in solid-state planar layouts ($\epsilon \gtrsim 1/D$), or as nonlinear optical aberrations (e.g., FOV vs. NA conflicts) and thermodynamic resource dilution ($I \propto P_{\mathrm{total}}/N$) in free-space atomic architectures.

\begin{figure*}[htbp!]
    \includegraphics[width=0.8\linewidth]{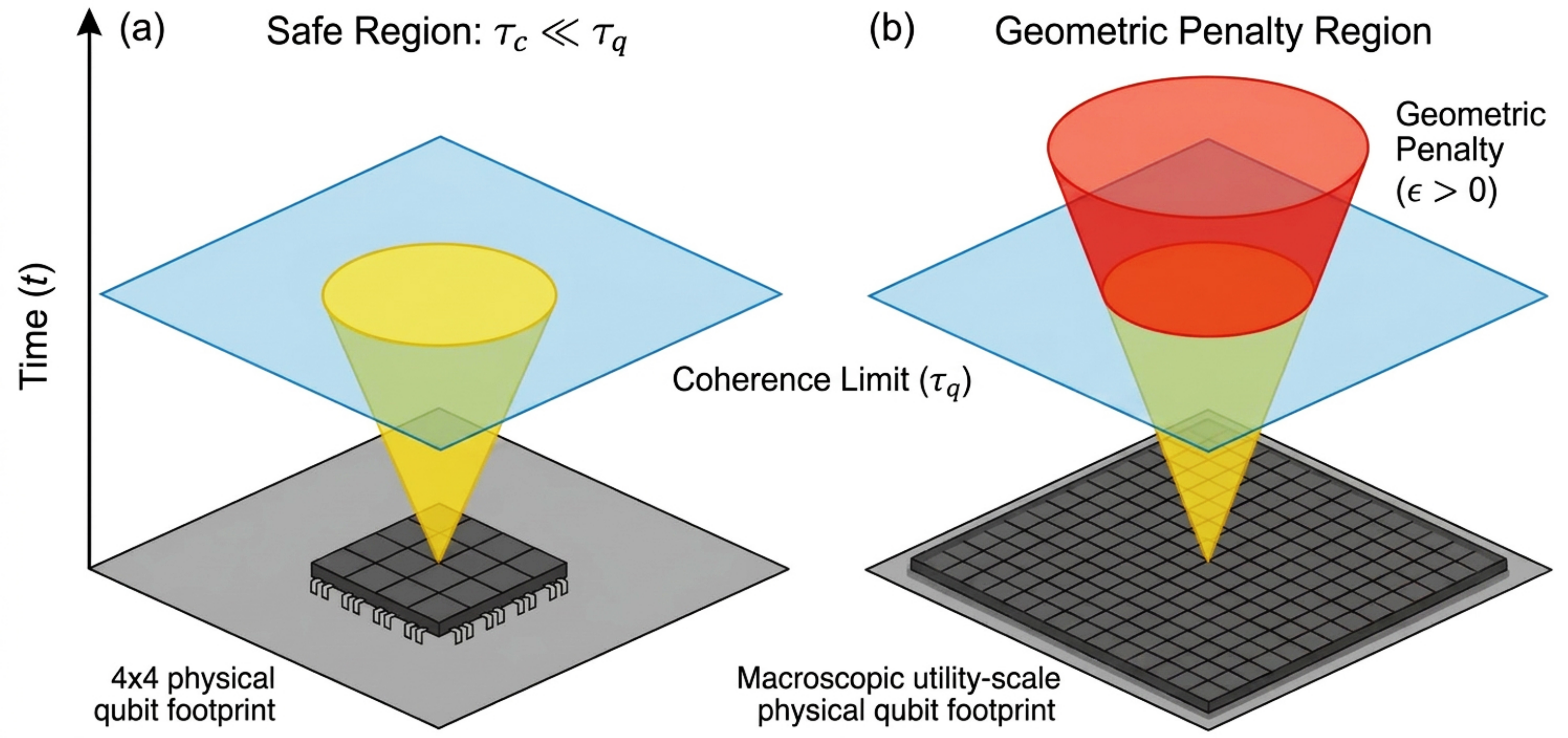}
    \caption{\label{fig1}\textbf{The Control Light Cone and geometric penalty in monolithic quantum architectures.} A 3D spacetime representation where the X-Y plane denotes the physical qubit footprint and the Z-axis denotes time. The blue horizontal plane represents the coherence lifetime limit ($\tau_q$). \textbf{(a)} At intermediate scales ($N < N_c$), the control signal cone (yellow) remains within the $\tau_q$ budget, enabling global coordination ($\tau_c \ll \tau_q$). \textbf{(b)} At utility scales ($N > N_c$), the spatial footprint expands such that the control cone breaches the $\tau_q$ horizon. The region exceeding the causal horizon (red shaded) incurs a superlinear geometric penalty ($\epsilon > 0$), physically mandating modular decomposition to truncate this penalty.}
    \end{figure*}

This scaling should be interpreted as an effective lower-bound baseline for coordination overhead under synchronization-constrained fault-tolerant cycles. Even with optimal asynchronous local decoders (e.g., Union-Find or sliding-window MWPM~\cite{dennis2002topological,delfosse2021almost}) and software-tracked Pauli frames~\cite{litinski2019game}, the causal requirement for adaptive non-Clifford feedforward across the macroscopic diameter $L$---which inherently relies on pre-distilled magic-state injection~\cite{bravyi2005universal}---preserves the $\Omega(N^{1/D})$ latency floor. Crucially, magic-state teleportation inevitably demands macroscopic classical coordination to update global Pauli frames~\cite{fowler2012surface}. This synchronization bottleneck persists regardless of local decoder speed, ensuring the geometric penalty $\epsilon > 0$ remains robust. Note that quantum error correction redistributes physical noise into overhead but does not eliminate this intrinsic space–time scaling penalty. This spacetime constraint is visualized in Figure \ref{fig1}.

Under synchronization-limited fault-tolerant execution, homogeneous scaling becomes asymptotically dominated by coordination:
          \begin{align}
          C_{\mathrm{hom}}(N) = \Omega(N^{1+\epsilon})
          \end{align}
Physically, $\epsilon$ is bounded from below by causal signal traversal: in any $D$-dimensional layout requiring global syndrome aggregation or adaptive feedforward, the classical coordination depth scales as $\Omega(N^{1/D})$. Under fixed QEC cycle normalization, this yields $C_{\mathrm{coord}}(N) \gtrsim \Theta(N^{1+1/D})$, establishing $\epsilon \gtrsim 1/D$ as a geometry-enforced baseline. Conversely, $\gamma$ is lower-bounded by network routing congestion; even under optimal expander topologies, $\gamma \ge 1$ holds as an information-theoretic limit on entanglement distribution complexity.

This superlinear bottleneck $C_{\mathrm{hom}}(N)$ represents a rigid geometric constraint imposed by finite signal propagation and classical feedforward latency. While future control co-design or non-local error correction schemes may mitigate the constant prefactor, the asymptotic exponent $\epsilon > 0$ remains strictly positive as long as global coordination relies on causal signal traversal.

In practical engineering terms, this theoretical bound manifests as severe diminishing returns. As the physical footprint $N$ increases, maintaining a constant global clock cycle necessitates prohibitively complex interconnect topologies and signal distribution networks, which rapidly violate physical limits regarding spatial crosstalk, routing congestion, and cryogenic heat dissipation. Conversely, if wiring density is capped to respect thermodynamic limits, the global clock cycle must be continuously slowed down to accommodate the increasing signal traversal distance. Consequently, long before the system reaches the utility scale required for fault tolerance, the global classical coordination latency ($\tau_c$) will inevitably approach or exceed the quantum coherence lifetime ($\tau_q$). Therefore, the $\epsilon$ penalty represents a hard architectural wall defined by spacetime causality, rather than a temporary fabrication hurdle.

\subsection{\label{sec2.3}Modular Decomposition and Interface Cost}
If the continuous spatial penalty $\epsilon$ is asymptotically prohibitive, the architectural imperative is to artificially bound the physical diameter of the synchronous system. This is the structural motivation for modular decomposition. By partitioning the architecture into bounded modules (with capacity $m \ll N_c$), the superlinear geometric penalty is effectively trapped within manageable local boundaries.

Within each module, isolation, control, and connectivity are optimized locally. Inter-module interactions are instead mediated through explicit interfaces (e.g., optical transduction via flying qubits). Through this architectural decoupling, the macroscopic scaling challenge is explicitly converted from an unsolvable continuous spatial problem (minimizing $\epsilon$) into a discrete, graph-theoretic network routing problem (optimizing $\gamma$). Under this modular paradigm, the total space–time cost becomes:
          \begin{align}
          C_{\mathrm{mod}}(N) = C_{\mathrm{module}}(N) + C_{\mathrm{interface}}(N)
          \end{align}

Because module size remains bounded, internal scaling is strictly linear:
          \begin{align}
          C_{\mathrm{module}}(N) = \Theta(N)
          \end{align}
Inter-module communication depends on network topology and the availability of active quantum repeaters. Without repeaters, transmission loss would introduce exponential overhead. Assuming repeater-enhanced links enabling polynomial routing, we model interface scaling generically as:
          \begin{align}
          C_{\mathrm{interface}}(N) = \frac{1}{\eta_{\mathrm{trans}}} \Theta(N^{\gamma}), \quad \gamma \ge 1
          \end{align}
where:
          \begin{itemize} 
          \item $\gamma$ encodes the macroscopic modular scaling exponent, strictly dictated by the network routing topology. The information-theoretic lower bound $\gamma \ge 1$ reflects that the space--time volume of communication cannot scale sublinearly with system size: even under ideal parallelization, the physical participation of $\Theta(N)$ distinct quantum carriers enforces a strictly linear cost floor.
          \item $\eta_{\mathrm{trans}} \in (0,1]$ denotes the dimensionless probability of quantum transduction efficiency \cite{awschalom2021development,lauk2020perspectives,kurpiers2018deterministic}.
          \end{itemize} 
Optimal topologies (e.g., hypercube or expander graphs) approach the limit $\gamma \approx 1$, while geometrically constrained networks yield $\gamma > 1$. Given the bounded internal module cost, the global modular asymptotic scaling is completely dominated by this routing complexity:
          \begin{align}
          C_{\mathrm{mod}}(N) = \Theta(N^{\gamma})
          \end{align}
          
\subsection{\label{sec2.4}Asymptotic Comparison and the Crossover Scale}

The architectural competition between homogeneous and modular reduces to exponent comparison:
          \begin{align}
          C_{\mathrm{hom}}(N) = \Omega(N^{1+\epsilon}), \quad C_{\mathrm{mod}}(N) = \Theta(N^{\gamma})
          \end{align}
          
Modularity becomes the structurally favored regime provided the geometric coordination penalty strictly exceeds the macroscopic modular scaling exponent:
          \begin{align}
          1 + \epsilon > \gamma
          \end{align}
This inequality formalizes a direct physical trade-off: modularity becomes favorable precisely when the geometric cost of macroscopic signal traversal ($\epsilon$) exceeds the topological overhead of routed entanglement distribution ($\gamma$). When this fundamental exponent inequality holds,
          \begin{align}
          \lim_{N \to \infty} \frac{C_{\mathrm{mod}}(N)}{C_{\mathrm{hom}}(N)} = 0
          \end{align}
          
Hence a finite crossover scale $N_c$ exists such that
          \begin{align}
          C_{\mathrm{mod}}(N) < C_{\mathrm{hom}}(N) \quad \text{for } N > N_c
          \end{align}
          
\textbf{Restoring Prefactors and Physical Interpretation}: To refine the crossover condition beyond asymptotic notation,  we restore the constant prefactors hidden in the asymptotic expressions:
          \begin{align}
          C_{\mathrm{hom}}(N) = A N^{1+\epsilon}, \quad  C_{\mathrm{mod}}(N) = \frac{B}{\eta_{\mathrm{trans}}} N^\gamma
          \end{align}
yielding
          \begin{align}
          N_c = \left( \frac{B/A}{\eta_{\mathrm{trans}}} \right)^{\frac{1}{(1+\epsilon)-\gamma}},
           \label{eq_Nc}    
          \end{align}
The existence of a finite, physically meaningful crossover scale ($N_c > 1$) is mathematically guaranteed if and only if this exponent is strictly positive. Because the elevated inter-module hardware costs and transduction inefficiencies guarantee a base $\frac{B/A}{\eta_{\mathrm{trans}}} \gg 1$, the necessary and sufficient condition for a phase transition is $(1+\epsilon)-\gamma > 0$. When this structural phase condition $1+\epsilon > \gamma$ holds, the superlinear geometric penalty inevitably overtakes the polynomial routing overheads at a finite scale.

Here, $A$ denotes the baseline cost of local physical interconnects in a monolithic architecture, while $B$ captures the fixed hardware overhead per inter-module interface (e.g., cryogenic isolation, optical coupling). The factor $\eta_{\mathrm{trans}}$ models operational inefficiency during entanglement generation and transfer, and is therefore treated as a stochastic resource overhead (expected retries) rather than a static hardware constant. Interface inefficiency ($\eta_{\mathrm{trans}} < 1$) and elevated inter-module costs ($B > A$) shift $N_c$ to larger scales but do not eliminate the existence of a crossover provided $1+\epsilon > \gamma$. This relation makes explicit that the crossover scale depends not only on asymptotic exponents, but also on the relative hardware cost structure.

The parameter regime considered here reflects a synthesis of results across fault-tolerant architectures and quantum networking. In surface-code-based systems, the quadratic scaling between physical qubits and logical code distance, combined with finite-speed decoding and feedforward, leads to a coordination overhead governed by an effective geometric exponent $\epsilon \approx 1/D$, where $D$ is the spatial dimensionality of the processor. This scaling assumes synchronization-limited fault-tolerant cycles, in which global coordination latency contributes directly to the effective space–time resource volume. Under current engineering constraints, thermodynamic limits and wiring density considerations effectively restrict large-scale monolithic implementations to $D \approx 2$, yielding a representative $\epsilon \approx 0.5$ for planar layouts \cite{fowler2012surface, google2023suppressing,tomita2014low}.

Crucially, this crossover scale is parameter-dependent rather than universal. For substrates with higher effective connectivity---such as dynamically reconfigurable neutral-atom arrays or 3D optical lattices---the geometric penalty is compressed (yielding a smaller effective $\epsilon$), while the relative interface cost $B/A$ increases due to the demanding photonic extraction hardware. These shifts displace the crossover boundary to larger scales (e.g., $N_c \sim 10^6$--$10^7$), but they do not alter the fundamental inequality $1+\epsilon > \gamma$. The asymptotic necessity of modularity therefore remains intact across all known physical platforms.

In modular architectures, repeater-assisted, non-planar connectivity (e.g., expander-like topologies) can approach near-linear scaling, corresponding to $\gamma \approx 1$ in the asymptotic limit \cite{baker2020time}. This exponent should be interpreted as an effective mean-field parameter that aggregates topological routing and physical entanglement distribution overhead under the assumption of sufficiently parallelized network operation.

Meanwhile, although current microwave-to-optical transduction efficiencies remain in the $10^{-3} - 10^{-2}$ range, a value $\eta_{\mathrm{trans}} \sim 10^{-1}$ is commonly identified as a near-term target threshold for utility-scale distributed quantum computing  \cite{awschalom2021development,lauk2020perspectives,kurpiers2018deterministic}. At the hardware level, this estimate reflects order-of-magnitude differences in footprint, cryogenic complexity, and integration overhead between on-chip interconnects and inter-module transduction interfaces. This justifies a baseline hardware cost ratio:
          \begin{align}
          \frac{B}{A} \sim 10 - 10^2.
          \end{align}

\begin{figure}[htbp!]
\includegraphics[width=0.95\linewidth]{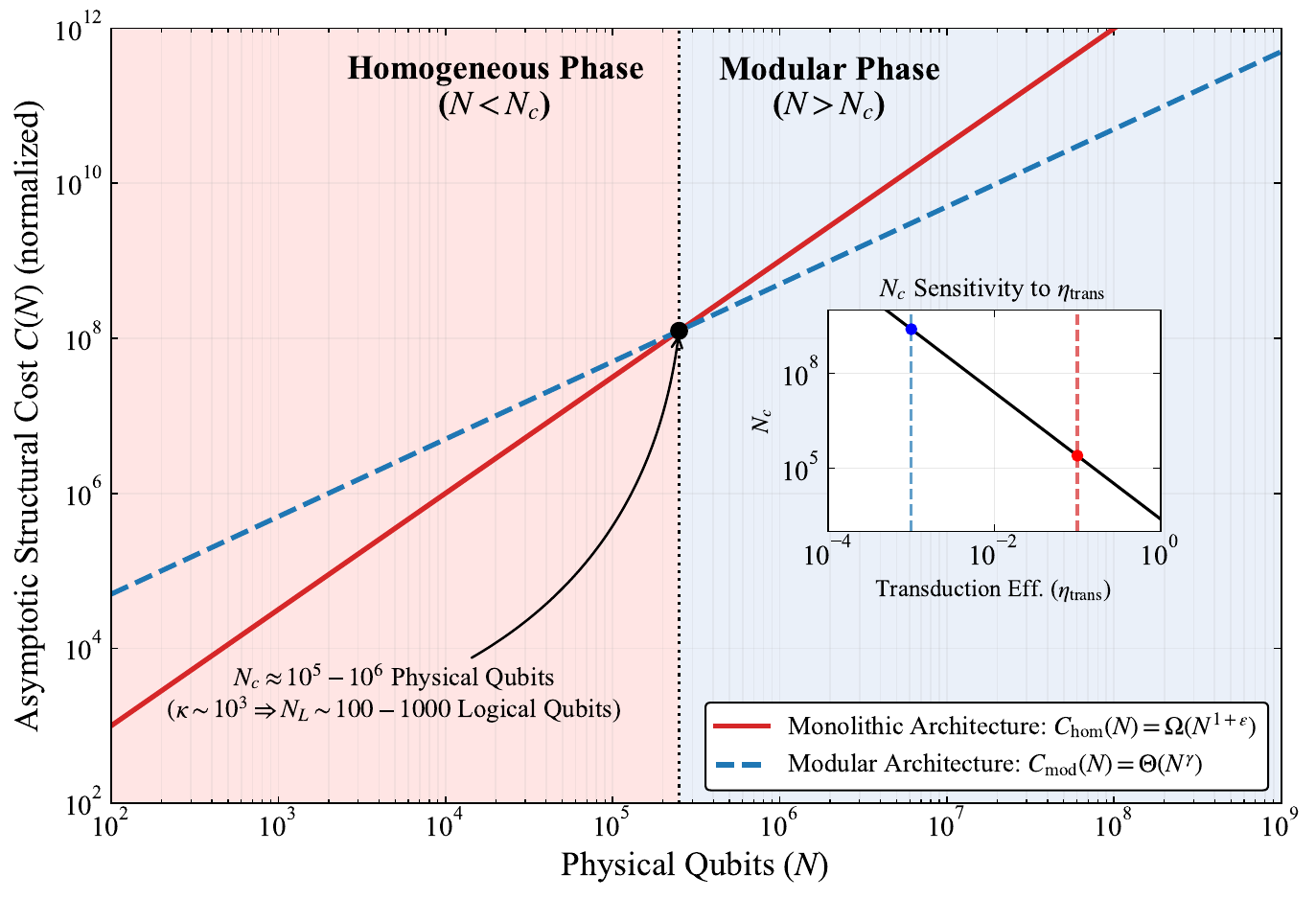}
\caption{\label{fig2}\textbf{Scaling-induced architectural phase transition and modular crossover.} 
Asymptotic structural cost $C(N)$ versus physical qubit count $N$, plotted on a log-log scale, 
where the physical footprint relates to logical capacity as $N \approx \kappa N_L$, with 
$\kappa \sim 10^3$ representing a typical surface-code overhead factor.
The homogeneous architecture (red solid line) incurs a superlinear coordination penalty, $C_{\mathrm{hom}}(N)=\Theta(N^{1+\epsilon})$ with $\epsilon>0$, arising from finite signal 
propagation and control bandwidth constraints;
the modular architecture (blue dashed line) achieves $C_{\mathrm{mod}}(N)=\Theta(N^\gamma)$, where $\gamma$ reflects network routing efficiency.
Their intersection defines a crossover scale $N_c(A, B, \eta_{\mathrm{trans}})$ see Eq.~(\ref{eq_Nc})), beyond which modular architectures become structurally favored ($C_{\mathrm{mod}} < C_{\mathrm{hom}}$).
Although the asymptotic competition is governed by the exponent inequality $1+\epsilon>\gamma$, the precise crossover scale depends additionally on prefactors: the relative hardware cost ratio $B/A$ (intermodule interfaces 
vs. local interconnects) and the transduction efficiency $\eta_{\mathrm{trans}}$.
(Inset) Sensitivity of $N_c$ to transduction efficiency $\eta_{\mathrm{trans}}$. Reducing $\eta_{\mathrm{trans}}$ from a near-term target ($\sim 10^{-1}$, red marker) to 
current experimental baselines ($\sim 10^{-3}$, blue marker) shifts the crossover by two to three orders of magnitude, from $\sim 10^5$--$10^6$ to $>10^8$ physical qubits.
Shaded regions denote the homogeneous phase ($N < N_c$) and modular phase ($N > N_c$).
Improved interface efficiency shifts the crossover to smaller scales. However, under known physical constraints, this does not eliminate the transition, establishing modularity as a scaling-induced structural tendency rather than a transient engineering choice.}
\end{figure}

Substituting these representative parameters into the crossover expression plausibly places the architectural phase transition in the regime of
          \begin{align}
          N_c \sim 10^5 - 10^6.
          \end{align}
From an engineering perspective, this numerical regime is profoundly significant. Under standard topological codes with a physical-to-logical overhead factor of $\kappa \sim 10^3$, a footprint of $10^5$ to $10^6$ physical qubits corresponds precisely to the $100$ to $1000$ logical qubits required to achieve early fault-tolerant utility (i.e., computations beyond brute-force classical simulation). This reveals a profound architectural convergence: the transition from the NISQ era to utility-scale Fault-Tolerant Quantum Computing (FTQC) temporally and physically coincides with the structural phase transition from monolithic to modular architectures. This intersection of scaling laws and the resulting structural transition are visually synthesized in Figure \ref{fig2}.

While the underlying model is asymptotic, this parameter-informed inversion provides a consistency check linking scaling exponents to experimentally relevant system sizes. These values should be interpreted as order-of-magnitude estimates rather than precise constants. The key conclusion is therefore robust: while physical prefactors ($A, B, \eta_{\mathrm{trans}}$) shift the location of $N_c$, the existence of a finite crossover is governed solely by the exponent inequality $1+\epsilon > \gamma$. This framework yields predictions broadly consistent with current technological roadmaps, reinforcing modularity as a scaling-induced structural tendency.

It is crucial to acknowledge the extreme sensitivity of this crossover scale. The projection $N_c \sim 10^5-10^6$ strictly depends on achieving the near-term target $\eta_{\mathrm{trans}} \sim 10^{-1}$. If transduction efficiency remains bottlenecked at current empirical levels ($\eta_{\mathrm{trans}} \sim 10^{-3}$)\cite{pompili2021realization}, the phase transition pushes exponentially outward to the $>10^8$ regime. Therefore, $N_c$ is not a static physical constant, but a dynamic structural boundary heavily dictated by the network interface efficiency. Under current $\eta_{\mathrm{trans}} \sim 10^{-3}$, the homogeneous phase extends beyond near-term utility scales, justifying continued monolithic integration efforts until transduction thresholds are met.

We note that even for asymptotically efficient non-local codes such as qLDPC~\cite{panteleev2022asymptotically}, whose experimental viability has recently been demonstrated~\cite{bravyi2024high}, their requirement for long-range expander-graph connectivity faces severe embedding congestion in planar monolithic substrates, structurally favoring distributed modular routing. As long as topological optimization outperforms brute-force geometric scaling, modularity emerges as the asymptotically favored structural regime.

Based on the projected near-term transduction targets ($\eta_{\mathrm{trans}} \sim 10^{-1}$) \cite{awschalom2021development,lauk2020perspectives,kurpiers2018deterministic} and utility-scale roadmap analyses \cite{kim2023evidence}, the emergence of this specific $N_c \sim 10^5–10^6$ threshold profoundly alters the field's technological trajectory. It indicates that the current race to fabricate the largest monolithic chip will face inevitable diminishing returns. As systems approach this regime, the primary developmental bottleneck shifts abruptly from the physical fidelity of individual qubits to the macroscopic transduction efficiency of inter-module interfaces. Consequently, a structural paradigm shift emerges as asymptotically favored: rather than exclusively scaling homogeneous substrates, architectural evolution increasingly biases the development of highly efficient quantum interconnects and classical causal backbones necessary to coherently stitch disparate modules together.

It is important to note that this structural transition is ultimately driven by the rigorous constraints of finite classical coordination resources. Detailed numerical derivations demonstrating that representative monolithic architectures encounter a severe causal bottleneck at $N \approx 8.3 \times 10^4$ are provided in Appendix \ref{appA}.

\subsection{\label{sec2.5}Scope and Architectural Assumptions}
To ensure the broad applicability of the proposed scaling framework across diverse quantum computing modalities, we explicitly define the scope and underlying architectural assumptions. Our theory applies to any utility-scale quantum system characterized by: (i) finite-speed classical coordination and bounded control bandwidth; (ii) non-zero synchronization overhead across spatially or logically separated registers; (iii) finite qubit coherence lifetimes ($\tau_q < \infty$); and (iv) the requirement for causal consistency in real-time feedforward and error correction.

While the framework does not presuppose a specific qubit modality, error-correction code, or fabrication platform, we acknowledge that different hardware implementations---ranging from monolithic superconducting lattices with fixed connectivity to reconfigurable neutral-atom arrays with flexible interaction graphs---will substantially alter the crossover scale ($N_c$) and the effective geometric exponent ($\epsilon$). For instance, architectures with high-degree connectivity or atom-transport capabilities may delay the onset of coordination dominance but remain asymptotically subject to the inherent temporal and causal constraints derived herein. We thus treat modularity not as a transient engineering choice, but as an asymptotically favored structural response to the scaling of classical coordination complexity.

\section{\label{sec3}Distributed and Modular Quantum Computing: From Global Circuits to Execution Protocols}
\subsection{\label{sec3.1}The Physical Origin of LOCC: Preserving Causal and Temporal Decoupling}

In conventional quantum information theory, Local Operations and Classical Communication (LOCC) is introduced as an abstract constraint describing spatially separated laboratories. Within utility-scale quantum architectures, however, this interpretation is insufficient. Once computation extends beyond the causal horizons dictated by finite-speed information propagation—encompassing both the quantum Lieb–Robinson limits and the classical control light cone—LOCC emerges as a physically necessary execution model.

A central architectural question is whether modular systems can maintain direct coherent quantum operations across macroscopic boundaries. In principle, deterministic non-local unitaries may be implemented via coherent interconnects, such as microwave buses, photonic links, or continuous-variable channels. However, when such operations are embedded within a fault-tolerant computational cycle, they impose effective synchronization constraints: participating modules must share a common logical clock domain, including aligned syndrome extraction, classical feedforward, and error-correction updates.

This synchronization constraint has a direct scaling consequence. The global coordination latency $\tau_c$ must span the system diameter $L \sim N^{1/D}$, thereby reintroducing the geometric coordination constraint identified in Section \ref{sec2.2}. As a result, the superlinear scaling characterized by the exponent $\epsilon > 0$ re-enters the critical execution path. Although the system is physically partitioned, it becomes causally equivalent to a monolithic architecture.

Avoiding this recurrence requires not only spatial separation, but causal and temporal decoupling. Modular subsystems must operate within independent clock domains, with coordination confined inside their respective control light cones. Crucially, even architectures employing coherent photonic links or continuous-variable interactions cannot avoid this constraint when embedded in fault-tolerant computation. Logical operations in error-corrected systems inevitably involve measurement-conditioned classical processing (e.g., syndrome extraction and decoding). Consequently, any coherent non-local operation that participates in a fault-tolerant cycle inherits synchronization constraints at the logical level, regardless of its physical implementation.

Therefore, the problem is not the presence of coherence, but the necessity of classical conditioning. Under finite-speed information propagation, any scalable execution model must decouple non-local correlation generation from the synchronous logical clock. This requirement leads directly to the LOCC paradigm. In this sense, LOCC should be understood not as a physical restriction, but as the operational manifestation of causal structure at scale.

\subsection{\label{sec3.2}Entanglement as an Asynchronous Resource}
Within this framework, entanglement assumes a qualitatively different operational role compared to circuit-based models. Rather than being generated on demand within a synchronized computational sequence, entanglement functions as a pre-distributed resource that enables asynchronous composition of non-local operations.

This distinction is essential for preserving causal decoupling. If entanglement generation were required during a logical cycle, it would introduce long-range temporal dependencies and re-couple distant modules. Instead, entanglement must be established outside the critical path through probabilistic generation, buffering, purification, and routing. Once established, entangled resource states serve as a temporal buffer.

Non-local logical operations are implemented via local measurements and classical communication conditioned on measurement outcomes. Because classical communication tolerates latency and does not require phase coherence, these operations do not impose global synchronization at the quantum level. From a systems perspective, entanglement thus acts as an asynchronous resource reservoir, separating the slow timescale of macroscopic correlation distribution from the fast timescale of local fault-tolerant execution. This separation is the key mechanism enabling modular architectures to avoid the geometric coordination penalty.

Critically, this asynchronous supply chain explicitly breaks the global clock synchronization barrier inherent in monolithic fault-tolerant cycles. By decoupling entanglement provisioning from the logical timeline, each module advances its local QEC cycles independently, preventing the slowest network link from stalling the global computational frontier.

\subsection{\label{sec3.3}Distributed Execution Model and Scaling Implications}
The adoption of an LOCC-based execution model profoundly alters the scaling behavior of large-scale quantum systems. In monolithic architectures, non-local operations directly contribute to coordination overhead, coupling logical depth to system diameter. In contrast, LOCC-based architectures shift this burden to the preparation and management of entanglement resources. This induces two structural changes.

First, the dominant scaling cost transitions from geometric coordination to network-mediated routing, consistent with the modular cost model introduced in Section \ref{sec2.3}. The effective scaling exponent $\gamma$ is determined by network topology and routing efficiency rather than spatial embedding. Second, the execution model becomes inherently layered. Local quantum processors perform high-fidelity operations within bounded regions, while a separate network layer manages entanglement generation and distribution. A classical control plane coordinates both layers through measurement outcomes and resource scheduling, without enforcing global synchronization.

This architecture replaces a rigid geometric constraint with a graph-theoretic one. As long as routing complexity grows sufficiently slowly (i.e., $\gamma < 1+\epsilon$), the modular architecture retains its asymptotic advantage. This structurally motivates the need for an execution model that natively supports such causal decoupling at scale.

\subsection{\label{sec3.4}Emergence of LOCC in the Large-Scale Limit}

The physical reasoning behind this structural necessity is straightforward. Any direct coherent non-local unitary acting across modules must be embedded within a fault-tolerant logical cycle. Such cycles inevitably inherit measurement-conditioned classical dependencies, imposing strict synchronization constraints across participating modules. This synchronization forces the global coordination latency $\tau_c$ to scale with system diameter, immediately reintroducing the superlinear coordination cost $\Omega(N^{1+\epsilon})$ identified in Section \ref{sec2.2}.

To avoid this geometric penalty, the architecture must explicitly decouple non-local correlation establishment from the synchronous execution timeline. Consequently, under known physical mechanisms, any scalable architecture operating beyond its control light cone is structurally biased toward an LOCC-equivalent execution model: long-range quantum correlations are pre-distributed asynchronously and consumed via local operations and classical communication. We emphasize that this argument does not exclude alternative physical realizations at small scales or in non-fault-tolerant regimes, but establishes the asymptotic constraint imposed by finite-speed information propagation.

Alternative paradigms, including photonic cluster-state or fusion-based quantum computing (FBQC) \cite{bartolucci2023fusion}, similarly reduce to LOCC-equivalent structures when embedded in fault-tolerant architectures. In these measurement-based frameworks, while deterministic unitaries are replaced by probabilistic fusions, the causal requirement to process measurement outcomes and update logical frames remains firmly bounded by classical coordination latency. This reinforces the universality of the $\tau_c \ll \tau_q$ constraint: even in ballistic photonic systems, the classical decoding backbone must finish within the temporal window defined by resource state availability or optical buffering limits.

\begin{figure*}[htbp!]
    \includegraphics[width=0.9\linewidth]{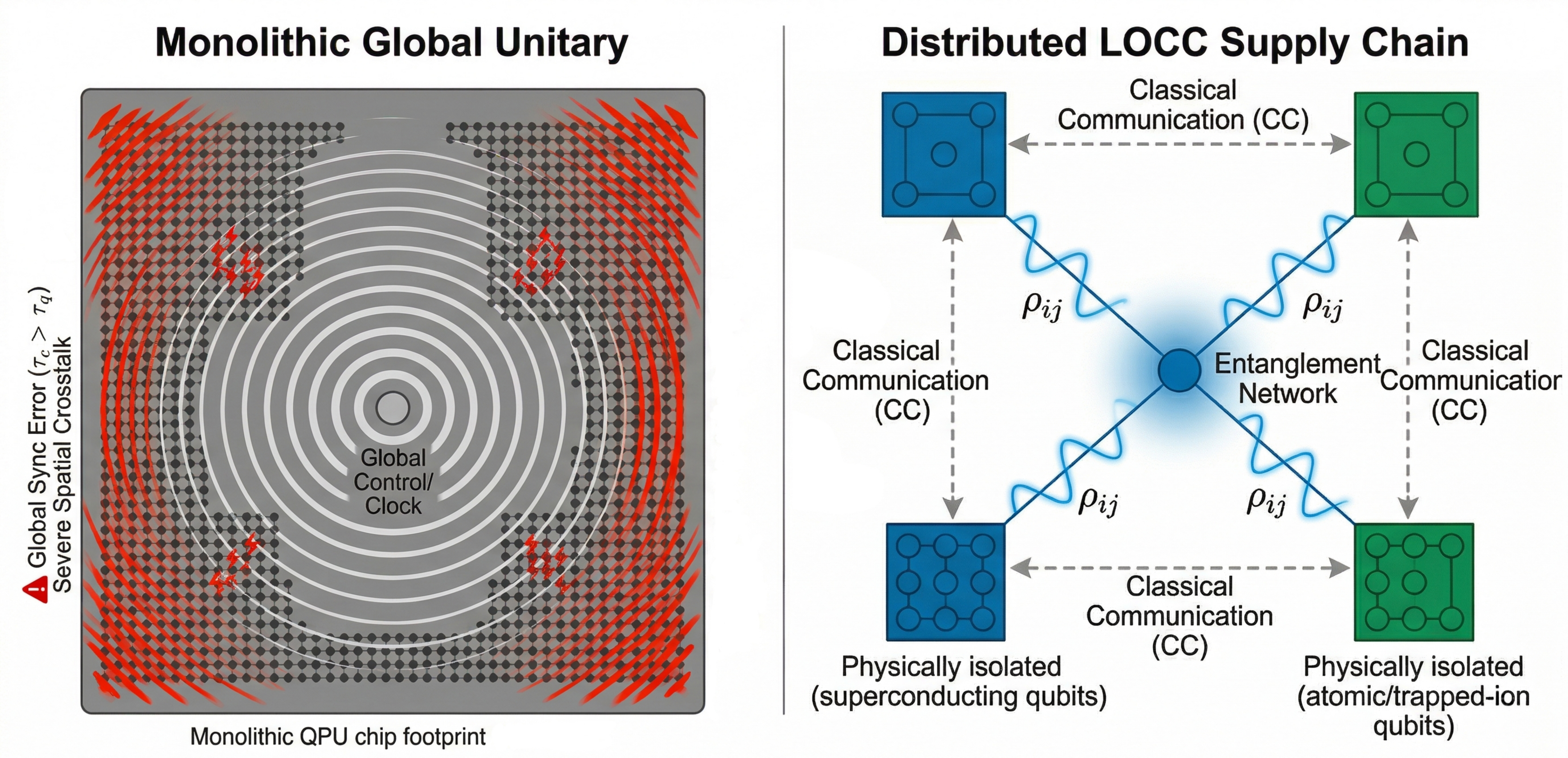}
    \caption{\label{fig3}\textbf{Conceptual shift from monolithic integration to a distributed LOCC supply chain.} \textbf{Left:} Monolithic global unitaries suffer from severe spatial crosstalk and global synchronization errors ($\tau_c > \tau_q$) as physical boundaries expand. \textbf{Right:} The distributed paradigm enforces strict modular isolation. Nonlocal operations are synthesized via Local Operations and Classical Communication (LOCC), consuming bipartite entanglement (managed as physical state $\rho_{ij}$ at the hardware layer, and abstracted as metadata tuples $e_{ij}$ for classical scheduling) pre-distributed asynchronously by a dedicated quantum network, thus accommodating hardware heterogeneity while circumventing the geometric penalty.}
    \end{figure*}

Consequently, beyond the crossover scale $N_c$, LOCC emerges as the canonical execution framework compatible with strict spatiotemporal causality under known physical mechanisms. It provides the operational mechanism for extending computation beyond the control light cone without collapsing back into monolithic behavior. This fundamental conceptual shift—from monolithic global coordination to a causally decoupled, distributed LOCC supply chain—is illustrated in Figure \ref{fig3}.

\section{\label{sec4}Quantum Networking: The Entanglement Supply Chain and Topological Routing}
\subsection{\label{sec4.1}The Operational Divergence: Asynchronous Pre-distribution vs. Data Transport}
The architectural mandate for modularity, established in Section \ref{sec3}, relies intrinsically on a network to bridge physically isolated subsystems. However, the operational paradigm of a quantum network diverges sharply from its classical counterpart. Classical networks rely on a "store-and-forward" model, dynamically buffering and routing encoded data packets across complex topologies.

Basic quantum mechanics prohibits this paradigm. The no-cloning theorem prevents the buffering and duplication of unknown quantum states, while measurement-induced collapse precludes non-destructive in-transit routing inspection. Attempting to directly transport active computational data across a network exposes fragile logical states to catastrophic loss and decoherence.

Consequently, the core operational function of a quantum network is not active data transport, but resource provisioning. As established by foundational quantum internet research \cite{wehner2018quantum}, the network must operate via asynchronous pre-distribution: it generates and distributes bipartite or multipartite entangled states in the background, decoupling the probabilistic, high-latency physical distribution process from the strict, deterministic timing constraints of the critical computational path. Pre-allocated entanglement decouples resource provisioning from the critical execution path. Our framework formally couples this supply chain to the $\tau_c \ll \tau_q$ temporal constraint, effectively circumventing end-to-end transmission delays during active logical cycles.

\subsection{\label{sec4.2}The Physical Interface and Transduction Efficiency ($\eta_{\mathrm{trans}}$)}
In a modular architecture combining distinct physical platforms, asynchronous pre-distribution must traverse severe physical barriers. Modules optimized for distinct axes of the design space often operate at vastly different energy scales—for instance, bridging the microwave frequencies of superconducting compute cores with the optical transitions of long-lived atomic memories. Crossing these domain boundaries requires active quantum transduction and entanglement distillation \cite{awschalom2021development,lauk2020perspectives,kurpiers2018deterministic}.

These physical processes are inherently lossy and probabilistic due to thermal noise, mode mismatch, and coupling inefficiencies. The aggregate success probability of establishing a high-fidelity entangled link across these boundaries is captured by the end-to-end physical transduction efficiency, $\eta_{\mathrm{trans}}$\cite{pompili2021realization}.

As established in the scaling model of Section \ref{sec2}, $\eta_{\mathrm{trans}}$ is not merely a component metric; it is the critical denominator dictating the crossover scale $N_c$. A low transduction efficiency drastically inflates the time and physical resources required to pre-distribute entanglement (scaling inversely with this probability as $\Theta(\eta_{\mathrm{trans}}^{-1})$), effectively throttling the entire computational supply chain. Elevating $\eta_{\mathrm{trans}}$ through improved electro-optic materials or high-coherence acoustic intermediaries is therefore a fundamental hardware prerequisite for rendering distributed architectures economically viable at utility scales.

\subsection{\label{sec4.3}Network Topology and the Routing Exponent ($\gamma$)}
Assuming a baseline transduction efficiency, the network must route entanglement across spatially distributed modules via entanglement swapping (e.g., Bell-state measurements at intermediate repeater nodes). Without active quantum repeaters, channel transmission loss scales exponentially with distance, causing scalable polynomial performance to collapse entirely.

Assuming repeater-enhanced links, the asymptotic cost of the network is dictated entirely by its graph-theoretic topology. This routing complexity directly dictates the macroscopic modular scaling exponent $\gamma$ in the interface cost model, $C_{\mathrm{interface}}(N) = (1/\eta_{\mathrm{trans}})\Theta(N^{\gamma})$. If modules are arranged in geometrically constrained, low-dimensional topologies (such as a 2D planar grid), the routing overhead and path congestion grow rapidly, resulting in $\gamma \geq 1$.

Conversely, highly connected topologies, such as expander graphs or hypercubes \cite{pant2019routing}, can logarithmically reduce path lengths and alleviate congestion, approaching the information-theoretic lower bound of $\gamma = 1$. The fundamental task of quantum network architecture is to optimize this topological routing structure. For modular architectures to asymptotically dominate monolithic scaling, the network design must guarantee that this scaling exponent ($\gamma$) remains strictly smaller than the total space–time scaling exponent ($1+\epsilon$) of the homogeneous substrate: $\gamma < 1+\epsilon$.

\subsection{\label{sec4.4}Resource Bookkeeping and the Timing Crisis}
The physical entanglement generated by the network layer must be made visible to higher architectural layers to be utilized in computation. To achieve this without violating isolation boundaries, the network layer abstracts the successfully pre-distributed entanglement into semantic metadata tuples:
          \begin{align}
          e_{ij} = (i, j, F, \tau_q^{(p)}, t_{\mathrm{gen}})
          \end{align}
where $i$ and $j$ denote the module endpoints, $F$ is the heralded fidelity, $\tau_q^{(p)}$ represents the strict high-percentile coherence deadline (e.g., ensuring $p = 99.9\%$ fidelity retention), and $t_{\mathrm{gen}}$ is the timestamp of generation.

Crucially, the quantum network only exposes this consumable interface; it does not dictate its usage. The responsibility to track, allocate, and schedule these resource tuples falls exclusively to the classical control plane. This division of labor exposes a fundamental timing crisis: unlike classical network links which possess indefinite memory, pre-distributed quantum entanglement has a strict expiration date governed by $\tau_q$.

If the classical control plane requires a routing and scheduling latency ($\tau_c$) that approaches or exceeds $\tau_q$, the resource tuple $e_{ij}$ will physically decohere before it can be committed to a logical operation. Therefore, to prevent the entanglement supply chain from collapsing under its own decoherence, the classical coordination infrastructure must be hyper-optimized to ensure $\tau_c \ll \tau_q$. This timing crisis requires the classical control plane to operate as a low-latency, authoritative coordinator for the distributed quantum infrastructure—a structural necessity formalized in Section \ref{sec5}.

\section{\label{sec5}Classical Control as a Structural Necessity: The Causal Backbone and Timing Constraints}
In large-scale quantum computing systems, the classical control plane is often mischaracterized as a temporary engineering scaffold that might eventually be replaced by "all-quantum" control logic. From a system-level perspective, this assumption deeply misunderstands the operational boundaries imposed by quantum mechanics. The dominance of classical control is not an artifact of current technological immaturity; it is a structural necessity derived directly from measurement-induced asymmetry and tightly bounded spacetime constraints.

\subsection{\label{sec5.1}Measurement-Induced Asymmetry and the Locus of Control}
Quantum states are inherently ephemeral, stateless execution units. The no-cloning theorem prohibits state duplication for concurrent branching (e.g., if-else pathways), while measurement-induced collapse precludes non-destructive routing evaluations. Consequently, quantum hardware cannot autonomously govern its own control flow or persistently store its execution history.

Therefore, quantum mechanics rigorously enforces a strict structural mandate: all control logic—including conditional branching, iterative looping, and state persistence—must explicitly reside within the classical plane. This fundamental asymmetry necessitates a strict operational separation. The classical domain—where information can be cloned, persistently stored, and deterministically evaluated—must serve as the sole locus of control. The classical control plane maintains the global execution state vector and ensures global state persistence. It acts as the causal backbone, driving the stateless quantum hardware through sequential, parameterized evolutions under the LOCC execution framework established in Section \ref{sec3}.

\subsection{\label{sec5.2}Resolving the Timing Crisis: The $\tau_c \ll \tau_q$ Imperative}
As the causal backbone, the classical control plane must orchestrate the entanglement supply chain introduced in Section \ref{sec4}. The quantum network provisions entanglement resources, tracked via the semantic metadata tuples previously defined as $e_{ij} = (i, j, F, \tau_q^{(p)}, t_{\mathrm{gen}})$. Crucially, these resources possess a strict physical expiration governed by the quantum coherence lifetime $\tau_q$.

The classical control plane is tasked with executing a computationally dense feedback loop: acquiring measurement outcomes (heralds or syndrome bits), decoding this information to make global routing or error-correction decisions, and actuating the subsequent feedforward control pulses. Let the total latency of this classical control loop be denoted as $\tau_c$. For adaptive distributed protocols and real-time quantum error correction to succeed \cite{kim2023evidence,fowler2012surface,google2023suppressing,lamport1978time}, the system architecture must guarantee the strict temporal inequality:
          \begin{align}
          \tau_c \ll \tau_q
          \end{align}
This inequality is not a flexible performance optimization metric; it is a hard precondition for logical correctness in feedback-driven operations. If $\tau_c \gtrsim \tau_q$, the pre-distributed entanglement or encoded logical state will physically decohere before the classical routing decision or correction pulse arrives, causing the execution protocol to fail.

This temporal imperative renders centralized, cloud-based control architectures unsuitable for real-time critical paths. To minimize signal flight time and processing latency, the classical control plane must be physically distributed alongside the quantum modules, often necessitating cryogenic co-integration or specialized low-latency FPGAs \cite{fu2018microarchitecture} to force the classical latency $\tau_c$ into the shrinking temporal margins of the control light cone. Recent hardware milestones in highly integrated, 3K-compatible Cryo-CMOS controllers~\cite{patra202019,xue2021cmos} empirically validate this architectural imperative, demonstrating that the critical control generation can be physically relocated across the thermal gradient to circumvent the macroscopic wiring bottleneck without sacrificing control fidelity.

\subsubsection{\label{sec5.2.1}The Causal Locality Bound on Real-Time Quantum Control}

This locality requirement yields a strict causal bound that sets a stringent physical limit on remote control paradigms. By enforcing $\tau_c < \tau_q^{(p)}$, we derive the maximum allowable physical distance $L_{\mathrm{ctrl}}^{\max}$ between the quantum substrate and its classical causal backbone:

    \begin{equation}
    L_{\mathrm{ctrl}}^{\max} \leq \frac{c}{2n} \left( \tau_q^{(p)} - \tau_{\mathrm{decode}} - \tau_{\mathrm{feedforward}} \right)
    \label{eq:locality_bound}
    \end{equation}
where $c/n$ denotes the effective signal propagation velocity in the cryogenic transmission medium. Assuming representative superconducting parameters, the raw coherence lifetime is $\tau_q \sim 100\,\mu\mathrm{s}$. However, assuming a standard exponential decoherence model, maintaining a target fault-tolerant fidelity retention $F_{\mathrm{target}}$ requires the control loop to complete within a much stricter deadline $\tau_q^{(p)} = -\tau_q \ln(F_{\mathrm{target}})$.

For a typical threshold requirement of $F_{\mathrm{target}} = 0.999$, this rigorously bounds the execution window to $\tau_q^{(p)} \approx 0.001 \tau_q \sim 100\,\mathrm{ns}$. Substituting this into Eq.~(\ref{eq:locality_bound}) alongside hardware-accelerated decoders, the theoretical upper bound on the classical control radius $L_{\mathrm{ctrl}}^{\max}$ rapidly contracts to the $\mathcal{O}(10^{-1})$ to $\mathcal{O}(10^0)$ meter range. This centimetric-to-metric scale confirms that real-time control logic must be physically co-located with the quantum substrate, rendering centralized cloud-mediated paradigms physically incompatible with fault-tolerant critical paths. Substituting state-of-the-art hardware parameters into this bound yields a concrete physical coordination wall. As detailed in the numerical case study in Appendix~\ref{appA}, for a distributed surface code architecture, this causal collapse occurs at $N \approx 8.3 \times 10^4$ physical qubits, strictly preceding any economic crossover.

\subsection{\label{sec5.3}Global Coordination Latency as the Dominant Origin of $\epsilon$}

This requirement for bounded classical latency provides the definitive physical explanation for the asymptotic scaling collapse modeled in Section~\ref{sec2}. Recalling the coordination bottleneck derived in Section~\ref{sec2.2}, we now explicitly ground the physical origin of $\epsilon$: in monolithic fault-tolerant cycles, classical feedforward signals and routing instructions must routinely traverse the macroscopic causal diameter of the system. Whether this bottleneck physically manifests as electromagnetic propagation across $L \sim N^{1/D}$ in dilution refrigerators, or as rigid acoustic transit delays scaling with optical aperture size in atomic arrays, it directly imposes an unavoidable temporal latency limit.

This temporal execution cost compounds with the macroscopic spatial footprint, rigorously isolating global classical coordination latency as the dominant physical source of the geometric penalty ($\epsilon > 0$), thereby rendering secondary engineering constraints (e.g., wiring density limits and cryogenic capacity) asymptotically subdominant.

Bounding the module capacity ($m$) rigidly confines the high-frequency, time-critical classical feedback loops ($\tau_c$) within local physical boundaries. The superlinear penalty $\epsilon$ is decisively truncated because the global classical control plane is relieved of high-frequency synchronization, only executing slower, low-frequency coordination across the inter-module interfaces.

\subsection{\label{sec5.4}Resource Tracking and the Protocol Precursor}
A final architectural challenge emerges at the semantic interface. The classical control plane acts as the global scheduler, tracking the live ledger of available entanglement tuples $e_{ij}$ generated by the underlying quantum network.

However, in a massively parallel distributed system, multiple execution threads will inevitably compete for the same nonlocal network resources. Because these quantum resources are ephemeral (strictly bound by $\tau_q$), standard classical mechanisms for resolving race conditions—such as indefinite blocking or thread queueing—are computationally catastrophic. If a computational thread dynamically locks an entangled link but is delayed by classical decoding or routing logic, the entangled resource will physically expire while sitting in the queue.

To ensure deterministic computation under these extreme temporal constraints, the system architecture cannot rely on ad-hoc or probabilistic scheduling. It requires a rigorous, layered semantic contract between the quantum network layer (the resource supplier) and the classical execution layer (the resource consumer). This critical necessity directly motivates the design of a Layered Semantic Architecture and the introduction of active Reserve--Commit Protocols, which we formally establish in Section \ref{sec6}.

\section{\label{sec6}A Layered Semantic Architecture and the Reserve--Commit Protocol}

The preceding sections have constrained the architectural design into a strict set of physical mandates: the macroscopic system must be modular to truncate the geometric penalty $\epsilon$ (Section \ref{sec3}); it must rely on a quantum network for asynchronous resource pre-distribution (Section \ref{sec4}); and it must be governed by a distributed classical control plane acting as the causal backbone under strict $\tau_c \ll \tau_q$ limits (Section \ref{sec5}). Fulfilling these antagonistic constraints simultaneously requires moving beyond ad-hoc hardware integration. It rigorously demands a formal Layered Semantic Architecture.

\subsection{\label{sec6.1}Architectural Inevitability: Isolating Incompatible Constraints}
Architectural layering is not a software preference but a structural imperative to isolate incompatible physical constraints. Consider the hardware heterogeneity introduced in Section \ref{sec2}. The underlying physical layer may comprise disparate technologies—such as superconducting circuits optimized for fast gate speeds (controllability) and neutral atom arrays optimized for long coherence times (isolation).

If the global classical control plane attempted to micro-manage these disparate substrates directly, the routing logic and calibration overhead would grow exponentially, immediately violating the $\tau_c \ll \tau_q$ temporal constraint. To prevent control plane overload, the architecture must enforce strict semantic boundaries. The lowest level, the Quantum Hardware Layer, must absorb the localized complexity of physical pulse generation and local calibration. Above it, the Entanglement Management Layer (Section \ref{sec4}) abstracts away the physics of transduction and swapping, exposing only the standardized entanglement metadata tuples $e_{ij}$. Finally, the Classical Control Plane operates purely on these semantic abstractions, scheduling global logic without needing to parse whether the underlying physical qubits are realized via microwave cavities or optical traps.

This layered isolation physically shields the upper algorithmic execution from the heterogeneous complexity of the underlying hardware, ensuring semantic stability across technology generations\cite{awschalom2021development,cross2022openqasm}.

\subsection{\label{sec6.2}Deconstructing the Nonlocal Gate: A Composite Transaction}
Within this layered architecture, operations that cross module boundaries can no longer be conceptualized as primitive quantum gates. A nonlocal operation, such as a distributed logical CNOT, must be strictly deconstructed into a composite transaction executed via LOCC. This distributed transaction essentially involves five sequential stages:
          \begin{enumerate}
          \item \textbf{Query}: Requesting the Entanglement Management Layer for a valid, pre-distributed resource tuple $e_{ij}$.
          \item \textbf{Local Entanglement}: Actuating local gates between the computational data qubits and the network interface qubits within each respective module.
          \item \textbf{Measurement}: Projectively measuring the interface qubits.
          \item \textbf{Coordination}: Transmitting the classical heralds (measurement outcomes) across the distributed control plane.
          \item \textbf{Feedforward}: Executing the conditional Pauli corrections on the target data qubits.
          \end{enumerate}
          
This composite sequence is extremely fragile. If any step stalls—due to network congestion or classical processing delays—the participating quantum data qubits are forced to idle. In the quantum domain, idling implies decoherence. Therefore, the architecture requires a rigorous protocol to ensure this sequence either executes to completion without interruption or is preemptively aborted before any active computational state is engaged.

\subsection{\label{sec6.3}The Reserve--Commit Protocol: Executing under $\tau_q$ Constraints}
To guarantee the execution of these composite transactions without violating temporal constraints, the classical control plane must implement a strictly time-aware Reserve--Commit Protocol. Classical distributed systems frequently employ atomic commit protocols, such as Two-Phase Commit (2PC) \cite{gray2005notes}, to ensure consensus across nodes during network failures. However, these consensus protocols are not optimized for strict temporal scheduling; they permit indefinite blocking, which in quantum architectures leads to irreversible failure via decoherence.

To circumvent this, our Reserve--Commit protocol prioritizes predictive temporal feasibility over classical fault tolerance, explicitly designing around the hard physical deadline $\tau_q$:
          \begin{itemize} 
          \item \textbf{Phase 1 (Reserve / Pre-validation)}: The control plane queries the resource ledger to lock an entanglement tuple $e_{ij}$ for a specific computational thread. Crucially, the scheduler performs a strict temporal pre-flight check. It calculates a bounded worst-case projected latency $\tau_{\mathrm{exec}}^{\ast}$ (e.g., a high-percentile statistical bound) to accommodate classical processing jitter and network fluctuations. If this projected completion time exceeds the required high-percentile coherence deadline $\tau_q^{(p)}$ (e.g., ensuring $p=99\%$ fidelity retention):
              \begin{align}
              t_{\mathrm{current}} + \tau_{\mathrm{exec}}^{\ast} > t_{\mathrm{gen}} + \tau_q^{(p)}
              \end{align}
          the protocol enforces a fail-fast mechanism. It immediately aborts the reservation, preventing the thread from initiating an invalid operation and avoiding systemic deadlocks. Because establishing an absolute worst-case execution time (WCET) is impractical under physical network jitter, relying on pessimistic bounds would trigger excessive ``false aborts.'' To address this, the pre-validation streamlines throughput by employing probabilistic elastic deadlines (e.g., $99.9^{\mathrm{th}}$ percentile latency bounds). This dynamically co-designs the strictness of $\tau_{\mathrm{exec}}^{\ast}$ with the QEC decoder's capacity to absorb the resulting erasure rate. The Reserve phase locks tuples atomically across all participating modules; if any node fails the temporal pre-check, the entire distributed reservation is rolled back before physical consumption begins.
          \item \textbf{Phase 2 (Commit / Execution)}: If the reservation succeeds and the temporal margin is mathematically secure, the control plane issues the execution command. The participating modules definitively consume $e_{ij}$ and execute the measurement and feedforward sequence seamlessly.
          \end{itemize} 

By enforcing this temporal pre-validation, the protocol achieves logically deterministic resource allocation. Unlike prior scheduling frameworks  \cite{siraichi2018,murali2019full} that optimize for gate count or compiler depth, our Reserve--Commit protocol optimizes for strict temporal feasibility under hard $\tau_q$ constraints. While physical noise (e.g., photon loss or gate errors) may still corrupt the operation, the Reserve--Commit protocol systematically prevents scheduling-induced decoherence caused by classical race conditions and resource starvation.

\begin{figure*}[htbp!]
    \includegraphics[width=0.95\linewidth]{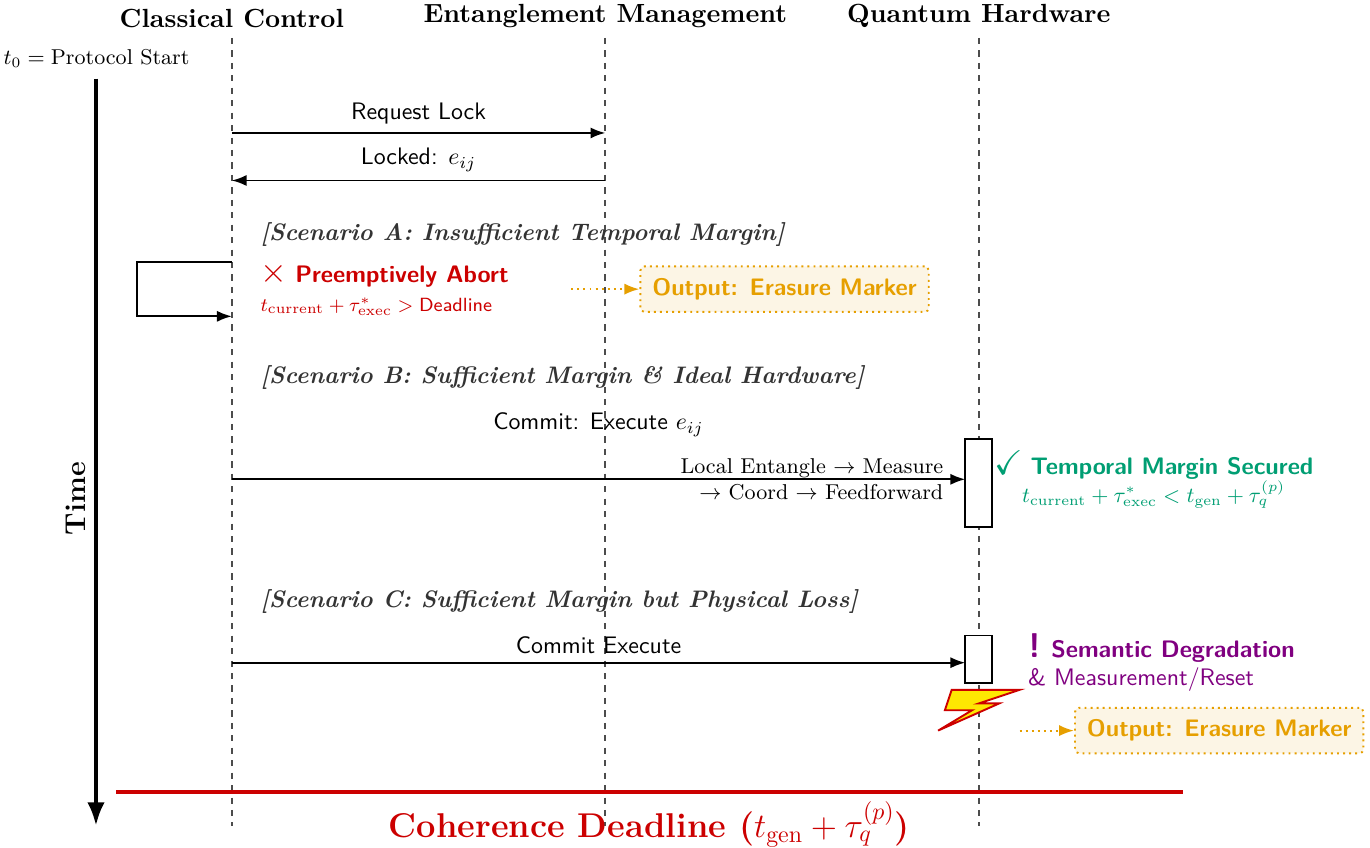}
    \caption{\label{fig4}\textbf{Time-aware Reserve–Commit Protocol timing diagram.} The protocol execution flow is shown across three architectural layers (Classical Control Plane, Entanglement Management Layer, and Quantum Hardware Layer), constrained by the physical coherence deadline ($t_{\mathrm{gen}} + \tau_q^{(p)}$, red line). In \textbf{Scenario A} (insufficient temporal margin), the control plane performs a preemptive abort when $t_{\mathrm{current}} + \tau_{\mathrm{exec}}^* > t_{\mathrm{gen}} + \tau_q^{(p)}$, explicitly outputting a semantic erasure marker (orange) to prevent scheduling-induced decoherence. In \textbf{Scenario B} (sufficient margin), the protocol commits to the execution of the LOCC sequence (Local Entangle $\rightarrow$ Measure $\rightarrow$ Coord $\rightarrow$ Feedforward) under the condition $t_{\mathrm{current}} + \tau_{\mathrm{exec}}^* < t_{\mathrm{gen}} + \tau_q^{(p)}$. In \textbf{Scenario C} (physical fault despite sufficient margin), the closed window for proactive erasure conversion necessitates semantic degradation (immediate measurement or reset) and classical Pauli frame updates.}
    \end{figure*}

In preemptively aborting transactions that exceed the coherence deadline, the protocol functions as a semantic classifier at the architecture layer: it flags the affected logical qubits as location-known erasure candidates. This metadata enables downstream QEC decoders to treat these events as erasure channels rather than uncharacterized depolarizing noise  \cite{tuckett2018ultrahigh,ma2023high,wu2022erasure,baranes2026leveraging}, without altering the underlying physical decoherence processes. The interaction between these semantic layers and the strict timing deadlines is detailed in Figure \ref{fig4}.

\subsection{\label{sec6.4}Ephemeral Rollbacks and Error Semantics}
The implementation of transactional protocols in quantum architectures exposes a final fundamental divergence from classical systems regarding error recovery. When a classical transaction aborts during the Commit phase, the system can reliably "rollback" the database to its pre-transaction state. Due to the no-cloning theorem and the destructive nature of intermediate measurements in LOCC protocols, quantum rollbacks are physically forbidden.

If a nonlocal transaction fails during the Commit phase—perhaps due to an unexpected detector dark count or a herald loss in transit—the original quantum information is irreversibly collapsed and cannot be restored. Consequently, the Reserve--Commit protocol must employ semantic degradation rather than state recovery. Upon transaction timeout, the protocol must actuate immediate projective measurements or fast resets. Forcing an active data qubit into a known state definitively severs it from the logical code block.

To prevent this from propagating as an unheralded multi-qubit error, the control plane immediately reports the precise spatiotemporal coordinates of the aborted qubit to the QEC decoder. Erasure-aware decoders treat such explicitly flagged, deterministically removed qubits equivalently to physical leakage or atom-loss events, dynamically adjusting the stabilizer parity checks around the known missing node \cite{tuckett2018ultrahigh,wu2022erasure}. Consequently, the risk of silent Pauli errors is strictly bounded to the brief waiting window $\tau_{\mathrm{exec}}^*$, rigorously justifying the semantic mapping to the higher erasure threshold. This rigorous reporting contract is what validates the layered architecture: it explicitly forces the responsibility of state recovery upwards, isolating it entirely within the realm of Quantum Error Correction (QEC) operating at the logical software layer.

\subsection{\label{sec6.5}Summary: Architectural Overheads and Bottlenecks}
The layered semantic architecture and the predictive Reserve--Commit protocol together provide the necessary operational structure to manage distributed quantum resources under strict physical constraints. However, this structure is not free. Layer abstraction introduces communication overhead, protocol handshakes consume critical fractions of the coherence time ($\tau_q$), and preemptive resource reservation may lead to hardware underutilization if fail-fast triggers activate too frequently. These architectural costs constitute a new class of systemic bottlenecks that constrain performance independently of baseline hardware fidelity. The following section analyzes these architectural bottlenecks, quantifying the trade-offs between semantic stability and operational efficiency to determine the true scalability limits of the proposed modular framework.

\section{\label{sec7}Architectural Bottlenecks and the Imperative for Logical Fault Tolerance}

The layered semantic architecture and the Reserve--Commit protocol established in Section 6 provide the necessary determinism to manage a massive, distributed quantum system. However, this semantic stability is not physically free. The mechanisms designed to shield the logical layer from hardware heterogeneity inevitably introduce structural overhead. To ensure the asymptotic validity of this modular approach, we must quantify these architectural bottlenecks and demonstrate how they structurally mandate the deployment of Quantum Error Correction (QEC).

\subsection{\label{sec7.1}Protocol-Induced Latency: The Semantic Overhead}
The most immediate consequence of a layered architecture is the accumulation of protocol-induced latency ($\tau_p$). In a monolithic setup, a classical feedback loop might execute as a direct hardware trigger. Conversely, in our layered architecture, nonlocal LOCC operations require explicit interface transactions: the execution layer must sequentially request resources from the network layer, perform the temporal pre-flight checks of the Reserve phase, and await inter-module heralds before issuing the Commit signal.

This classical handshake latency directly squeezes the usable quantum coherence window. The effective computing time for the quantum data qubits is reduced to:
          \begin{align}
          \tau_{\mathrm{compute}} = \tau_q - (\tau_c + \tau_p)
          \end{align}

Crucially, although this semantic overhead reduces the physical fidelity within the local coherence window, it structurally preserves the macroscopic scaling advantage derived in Section \ref{sec2}. As long as the protocol latency $\tau_p$ is strictly bounded (e.g., $O(1)$ or $O(\log N)$) and avoids superlinear growth, the fundamental exponent condition $\gamma<1+\epsilon$ remains robust. The architecture essentially pays a manageable temporal tax to eliminate the prohibitive superlinear spatial penalty of the monolithic substrate.

\subsection{\label{sec7.2}Resource Starvation and the Fail-Fast Penalty}

While protocol latency compresses the execution window, the most severe operational bottleneck in early-stage modular scaling is resource starvation. The Reserve--Commit protocol deliberately employs a fail-fast mechanism to prevent scheduling-induced deadlocks when physical entanglement cannot be provisioned before the strict temporal deadline ($t_{\mathrm{current}} + \tau_{\mathrm{exec}}^{\ast} > t_{\mathrm{gen}} + \tau_q^{(p)}$).

This protective mechanism directly couples macroscopic system utilization to the physical transduction efficiency ($\eta_{\mathrm{trans}}$) defined in Section \ref{sec4}. If the rate of entanglement generation across modular boundaries significantly trails the rate of logical consumption, computational threads will face chronic resource starvation. Consequently, the fail-fast condition will trigger frequently, forcing the system to abort and explicitly retry composite transactions. This high abort rate leads to systemic idling, where precious quantum data qubits inevitably decohere while waiting for valid entanglement tuples. Therefore, pushing the physical limits of $\eta_{\mathrm{trans}}$ is not merely a component-level optimization; it is the absolute physical prerequisite for maintaining high logical duty cycles in any distributed execution model.

\subsection{\label{sec7.3}Semantic Degradation as Structured Error Information}
When the fail-fast mechanism aborts a transaction, or when the Commit phase fails due to heralded physical loss (e.g., detector dark counts or photon loss), the original active quantum state is irreversibly corrupted. Consequently, the architecture must respond with semantic degradation.

\begin{figure}[htbp!]
\includegraphics[width=0.8\linewidth]{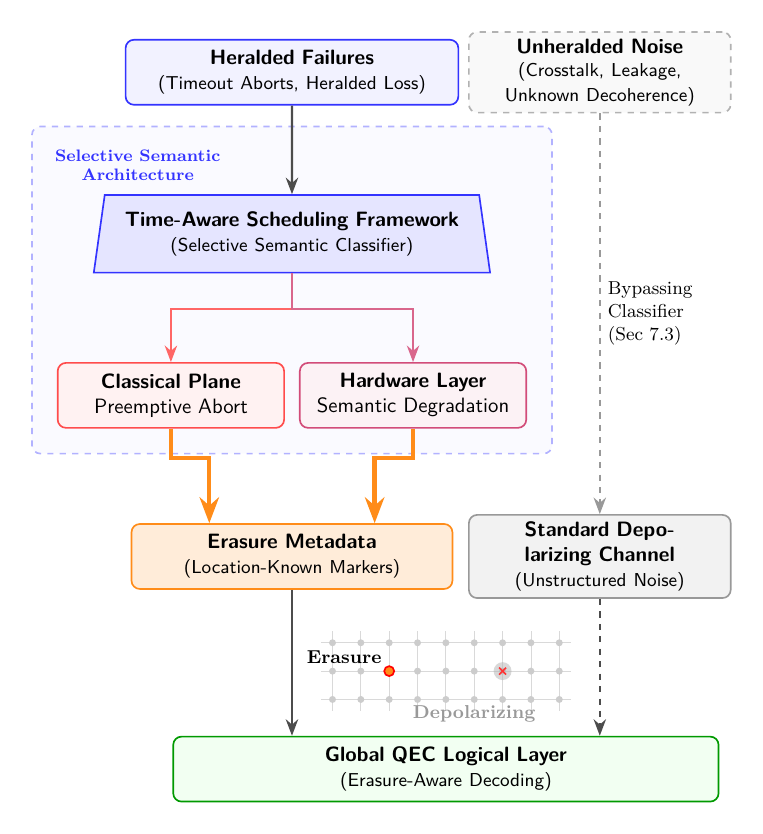}
\caption{\label{fig5}\textbf{The architectural semantic error degradation pipeline.} The time-aware scheduling framework operates as a selective semantic classifier: heralded protocol failures (timeout aborts, heralded photon loss) are converted into location-known erasure metadata, while unheralded local decoherence bypasses the classifier as standard depolarizing noise. This structured routing supplies downstream erasure-aware QEC decoders with explicit error metadata, elevating the effective fault-tolerant threshold and reducing syndrome decoding complexity.}
\end{figure}

From a system-level perspective, this degradation yields a structural advantage by functioning as a selective semantic classifier. Instead of allowing protocol-induced failures to manifest as uncharacterized depolarizing noise, the time-aware framework intercepts heralded failures (e.g., timeout aborts or heralded photon loss) and routes them into explicit, location-known erasure metadata. Recent theoretical frameworks and experimental validations corroborate this approach, demonstrating that converting physical failures into explicit erasure markers drastically reduces the complexity of subsequent syndrome decoding \cite{tuckett2018ultrahigh,ma2023high,wu2022erasure,baranes2026leveraging}. Conversely, unheralded physical noise is identified as an irreducible background that bypasses the classifier, preserving the physical integrity of the QEC depolarizing noise model. The role of the architecture as a semantic classifier is visualized in the error degradation pipeline of Figure \ref{fig5}.

Crucially, this architectural semantic classification is not merely a bookkeeping convenience; it directly relaxes the physical fidelity requirements for the hardware substrates. Standard surface code thresholds under uncharacterized depolarizing noise reside near $p_{\mathrm{th}} \approx 0.94\%$. However, recent theoretical results demonstrate that when a dominant fraction of these errors are explicitly flagged as location-known erasures, the fault-tolerant threshold elevates significantly toward the erasure regime ($p_{\mathrm{th}}^{\mathrm{erase}} \approx 3.1\% \text{--} 4.15\%$)~\cite{tuckett2018ultrahigh,wu2022erasure,baranes2026leveraging}. By intercepting out-of-bounds delays before they manifest as uncharacterized logical depolarizing noise, the Reserve--Commit protocol effectively converts protocol-induced failures into these high-threshold erasure channels. This provides quantifiable architectural headroom, allowing near-term quantum modules to operate at physically achievable fidelities while still sustaining macroscopic logical fault tolerance.

\subsection{\label{sec7.4}The Imperative for Logical Fault Tolerance (QEC)}
This structured error exposure ultimately defines the strict boundary between the physical architecture and the algorithmic workload. The physical hardware, the network supply chain, and the layered protocol stack can only isolate, schedule, and flag quantum errors; inherently constrained by the no-cloning theorem and measurement collapse, they cannot independently correct them.

Therefore, Quantum Error Correction (QEC) cannot be viewed as an algorithmic afterthought or a distinct software layer. It is the structural imperative of the entire architecture. The QEC logical layer acts as the ultimate macroscopic consumer of the distributed system: it absorbs the protocol-induced latency ($\tau_p$), tolerates resource starvation aborts via inherent retry logic, and actively consumes the structural erasure markers provided by the Reserve--Commit protocol to efficiently decode the global syndrome graph. Only through the deployment of this global logical layer can the localized physical subsystems and the severe temporal constraints of the distributed substrate be fully synthesized, finally presenting to the user the macroscopic abstraction of a fault-tolerant, universal quantum Turing machine. The following conclusion summarizes the implications of this integrated framework for the future trajectory of quantum computing infrastructure.

\section{\label{sec8}Conclusion and Future Outlook: The Dawn of Quantum Systems Engineering}

The framework developed in this perspective yields four central implications for the future of quantum systems engineering.

\textbf{Architectural Imperative.}
This perspective has demonstrated that the transition to utility-scale, fault-tolerant quantum computing is increasingly subject to progressive scaling pressure from strict spatial and temporal physics. When a monolithic quantum processor expands beyond its control light cone ($\tau_c \gtrsim \tau_q$), it incurs a superlinear geometric penalty ($\epsilon$) that subjects the system to progressive scaling pressure at utility scale. To truncate this penalty, the macroscopic system must modularize. Managing this distributed hardware demands an asynchronous entanglement network, a distributed classical causal backbone, and a rigorous Reserve--Commit protocol. Rather than a set of subjective engineering preferences, this layered architecture is the structurally necessary mechanism to translate the fragility of quantum mechanics into a macroscopic fault-tolerant architecture, ultimately allowing Quantum Error Correction (QEC) to efficiently absorb protocol-induced semantic erasures.

\textbf{The Mandate for Interoperable Co-Design.}
Because physical constraints across these layers are tightly coupled—such as the physical transduction efficiency ($\eta_{\mathrm{trans}}$) directly throttling resource availability, and the quantum coherence limit ($\tau_q$) strictly dictating the classical latency budget ($\tau_c$)—single-discipline optimization is no longer viable. Device physicists cannot optimize physical fidelities while ignoring transduction rates; network engineers cannot maximize throughput without minimizing protocol handshake latencies; and QEC theorists cannot assume perfect, location-agnostic noise models. To break these disciplinary silos, the community must establish a common interoperability framework. The layered reference architecture proposed herein serves precisely as this connective tissue, enabling cross-stack, hardware-software co-design without violating underlying physical laws.

\textbf{Quantitative Boundaries and Reference Blueprints.}
To navigate this technological transition, our theoretical framework yields specific, falsifiable boundaries for future system design. Under known physical mechanisms, we project that the architectural crossover point—where modularity definitively outperforms monolithic scaling—will occur at a macroscopic scale of $N_c \sim 10^5$–$10^6$ physical qubits, contingent on achieving the critical inter-module entanglement generation threshold $\eta_{\mathrm{trans}} \gtrsim 0.1$ required to prevent protocol-induced resource starvation. As derived in Appendix~\ref{appA}, the physical coordination wall strictly precedes this economic crossover $N_c$, confirming that modularity is a prerequisite for system survival before it becomes a resource-efficiency preference. This projection is not a distant theoretical horizon, but the immediate next decision node confronting current technological roadmaps (e.g., the targeted transitions toward 10,000-qubit modular systems in IBM's Kookaburra or advanced neutral-atom array proposals). While the specific entanglement tuples ($e_{ij}$), Reserve--Commit primitives, and erasure markers defined in this work are not the singular solution, they provide a crucial reference blueprint for operating at these macroscopic scales.
\textbf{Semantic Prerequisites for Future Interfaces.}
Ultimately, realizing a universal quantum computer requires evolving current instruction set architectures (ISAs) and cloud APIs into spatio-temporal aware interfaces. Future quantum interfaces must move beyond pure logical expressions and explicitly expose strict semantic prerequisites to the compiler: the error location (to facilitate deterministic erasure conversion), the statistical coherence deadline ($\tau_q^{(p)}$), and the operation atomicity (transactional guarantees). When hardware iterations finally converge beneath these robust semantic abstractions, the construction of utility-scale, fault-tolerant quantum computers will decisively shift from the chalkboards of theoretical physics to the pipelines of rigorous quantum systems engineering.

\appendix 
\section{\label{appA}Numerical Illustration: The Imminent Coordination Cliff}

To illustrate that the crossover $N_c \sim 10^5$ represents an imminent engineering boundary rather than an asymptotic abstraction, consider a representative monolithic superconducting array. Assume a distance-$d=31$ surface code ---a target regime motivated by recent experimental milestones in exponential error suppression~\cite{google2025quantum}, for which a logical error rate $\sim 10^{-6}$ per cycle is projected--- with a physical gate cycle $t_{\mathrm{cycle}}=1\,\mu\mathrm{s}$ and coherence window $\tau_q \approx 100\,\mu\mathrm{s}$. A hardware-accelerated Union-Find decoder incurs a baseline processing latency $\tau_{\mathrm{decode}} \approx 2.5\,\mu\mathrm{s}$~\cite{delfosse2021almost}.

The total global coordination latency scales as $\tau_c(N) = \tau_{\mathrm{decode}} + \tau_{\mathrm{ff}} + \alpha \sqrt{N} \cdot \tau_{\mathrm{route}}$, where $\tau_{\mathrm{ff}} \approx 0.5\,\mu\mathrm{s}$ captures feedforward overhead. To ground this scaling in current control hardware benchmarks, we note that state-of-the-art FPGA-based feedback systems achieve latencies as low as 110 ns~\cite{salathe2018low}, dedicated neural-network accelerators for quantum control exhibit 175 ns inference latency~\cite{xu2022neural}, deterministic inter-FPGA communication in distributed control has been demonstrated at 361.60 ns~\cite{oliveira2023fpga}, and a multi-FPGA architecture for lattice-surgery-based error correction has achieved sub-microsecond decoding latency~\cite{liyanage2024multi}.

Anchored by these baselines, we adopt a planar routing elongation factor $\alpha \approx \sqrt{2}$ (representative of Manhattan-style grid routing) and an effective control-distribution latency per lattice unit $\tau_{\mathrm{route}} \approx 115\,\mathrm{ns}$. Crucially, $\tau_{\mathrm{route}}$ is dominated by classical digital overheads---including FPGA-to-FPGA switch hops, multiplexing, and signal fan-out---rather than pure electromagnetic propagation.

Solving for the critical boundary where cumulative classical latency breaches a conservative $50\%$ safety margin of the coherence budget ($\tau_c(N) \approx 0.5 \tau_q = 50\,\mu\mathrm{s}$), we obtain $50\,\mu\mathrm{s} \approx 3.0\,\mu\mathrm{s} + 0.163\,\mu\mathrm{s} \times \sqrt{N}$, yielding a coordination wall at $N \approx 8.3 \times 10^4$ physical qubits. Varying $\tau_{\mathrm{route}}$ within a realistic hardware range ($80$--$150\,\mathrm{ns}$) shifts this boundary between $4.9 \times 10^4$ and $1.7 \times 10^5$, preserving the order-of-magnitude conclusion. Beyond this regime, the architecture enters the superlinear penalty phase ($\epsilon>0$) dominated by geometric routing latency rather than decoder throughput. This quantitative bound confirms that the structural transition toward modularity is physically mandated well before reaching the $10^6$-qubit utility scale.

\label{appc}
\bibliographystyle{apsrev4-2}  
\bibliography{apssamp}

\end{document}